\documentclass[12pt,a4paper]{article}
\usepackage{graphicx}

\begin{document}

\begin{center}
{\bf \Large Social phase transition in Solomon network}\\[5mm]

{\em 
{\large K. Malarz}\\[3mm]
Department of Applied Computer Science, Faculty of Physics and Nuclear Techniques, University of Mining and Metallurgy (AGH)\\
al. Mickiewicza 30, PL-30059 Krak\'ow, Poland.\\[3mm]
\today
}
\end{center}

\begin{abstract}
In this paper the Solomon network is simulated by means of 1D and 2D Ising model with additional ---  not only geometrical --- neighbors. 
A ``social phase transition'' at a non-zero Curie-like temperature is observed, also in one dimension.
The critical exponent describing the behavior of the magnetization in the vicinity of the transition is also evaluated.
\end{abstract}


\section{Introduction}
In many sociological models, the behavior of each person is influenced by
the neighbors, or influences the neighbors.
However, the neighbors in the workplace are different from the neighbors at home.
Thus the workplace neighborhood and the home neighborhood can be simulated by using two lattices of $N$ sites each, the home lattice and the workplace lattice.
In the workplace lattice, the people are numbered consecutively from $i=1$ to $i=N$ with helical boundary conditions.
The same people also show up on the home lattice, but an different order which is a random permutation of the order on the workplace lattice.
Thus each person has exactly one place on the home 
lattice, and each site on the home lattice is occupied by exactly one person,
just as is the case for the workplace lattice.
But the same person occupies two entirely different sites on the two lattices.

Such a network of two lattices is called a Solomon network (SN) \cite{solomon} because each person is shared equally by the two lattices, just as in the biblical 
story of King Solomon.

Within each lattice we have the usual type of interaction (Ising model, n-vector
model, Sznajd model, ...); added to it is the interaction of each person with
its own image on the other lattice. Thus in a chain of people $i$ with nearest 
neighbor connection, the variables at site $i$ interacts with the variable at
sites $i-1$ and sites $i+1$ as well as with the neighbors of the site $P(i)$
of the other lattice, where $P$ is the permutation of the numbers $i = 1,2,
\dots, N.$

A simplified version of this two-lattice model uses only one lattice, and 
defines the neighborhood of each site as being the nearest neighbors plus one
randomly selected site anywhere in the lattice.
For a chain the neighbors of $i$ are thus $i-1, i+1, i+R$ where $R$ is a random distance; these neighbors can be compared with those of the honeycomb lattice: $i\pm 1, i\pm L$.
Therefore it cannot be excluded that ordering is found also in one dimension.
This simplified model is closer to small-world networks \cite{herrero,pekalski,hong}.

\section{Model}
Here, we consider interaction between individuals in the same manner like interacting spins in the Ising model.
The intensity of interaction among neighbors at both lattices is described by the coupling constant $J$.
When $J$ is positive, neighbors at sites $i$ and $j$ tend to have the same opinion or share the same interests ($S_i=S_j$).
The external magnetic filed $H$ is then global trend or world-wide fashion and forces spins to have the same sign as $H$.
Then the temperature $T$ plays the role of ``social temperature'' and describes a degree of randomness in the behavior of the members of the community.

Before the system evolution according to Metropolis algorithm is started, additional neighbors of the $i$-th spin are chosen.
For a give spin, the neighbors are fixed during simulation.
During the simulation for each site at the lattice (individual, member of society) the energy change $\Delta E=E_{new}-E_{old}$ associated with a possible spin flip (mind change) from $S_i$ into $-S_i$ is evaluated, where the energy is given by the Ising form:
\begin{equation}
E=-J\sum_{<i,j>} S_iS_j
\end{equation}
The sum in Eq. (1) goes over all neighbors pairs.
For a given temperature $T$ (expressed in $J/k_B$ units) a spin is flipped if and only if the thermal probability $\exp(-\Delta E/k_BT)$ is greater than a random number taken uniformly from the interval $[0,1]$.

We consider 1D and 2D Ising model with one (1N) or two (2N) additional randomly selected neighbors.
The simulation is performed on $N=10^6$ sites with helical boundary conditions and initially, at $t=0$, all spins are pointing up ($S_i=+1$ for $i=1,2,\dots, N$).
Later we calculate the magnetization $m=\sum_{i=1}^NS_i/N$.
The time is measured in Monte Carlo steps (MCS) and one MCS is completed when all $N$ spins at the lattice where investigated (in type-writer order) if they should flip or not.
The value of the ``social magnetization'' $m$ is averaged over the last $10^4$ MCS.

\section{Results}
\begin{figure}
\begin{center}
\includegraphics[angle=-90,width=.8\textwidth]{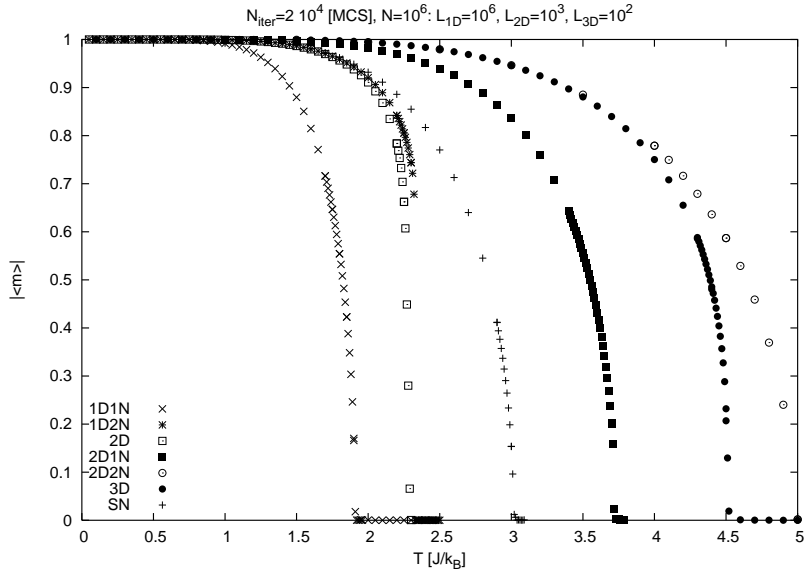}
\end{center}
\caption{The temperature dependence of the absolute value of the average magnetization.
Simulation takes $2\cdot 10^4$ [MCS].
The value of the magnetization is averaged over the last $10^4$ [MCS].
The systems contains one million sites.}
\label{fig_mag}
\end{figure}

It is well known from Ising himself that for the 1D case the Curie temperature $T_C=0$. Introducing one additional neighbor somewhere at the lattice in 1D case shifts the ``social Curie point'' towards $T_C\approx 1.9(1)$ (in units of $J/k_B$) as presented in Fig. \ref{fig_mag}.
This rough estimation of $T_C$ gives a little bit higher values than reported in \cite{pekalski,hong}.

\begin{figure}
\begin{center}
\includegraphics[angle=-90,width=.8\textwidth]{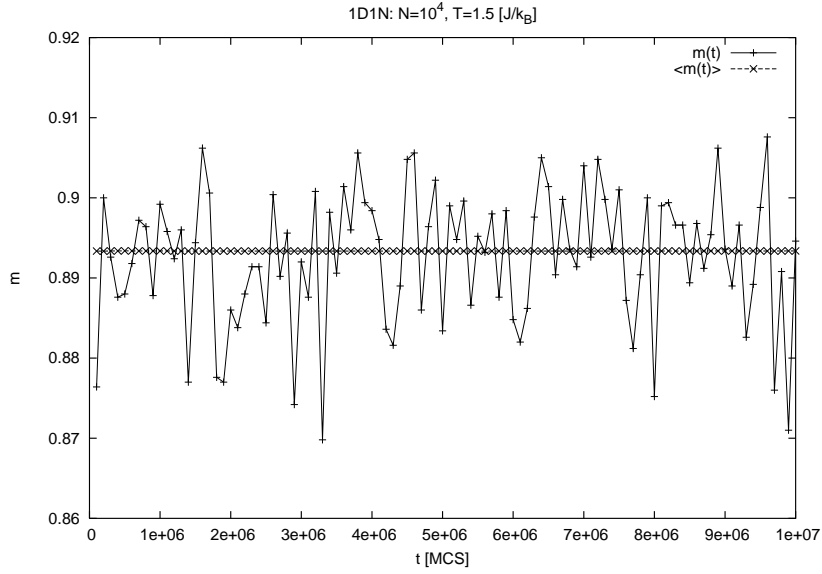}
\end{center}
\caption{Time evolution of magnetization $m(t)$ and its cumulative time average $\langle m(t)\rangle$ for 1D1N case and $T=1.5~[J/k_B]$.
The systems contains $N=10^4$ sites.}
\label{fig_time}
\end{figure}

For a given temperature $T$ the magnetization $m(t)$ and its cumulative time average $\langle m(\tau)\rangle=\sum_{t=1}^\tau \sum_{i=1}^N S_i/(N\tau)$ are stable except of small fluctuations (as presented in Fig. \ref{fig_time}).

The original SN model of two interacting Ising chains gives higher $T_C$ than the 1D2N case with two additional and random neighbors (see Fig. \ref{fig_mag}). 
The difference between these two models is that in the latter case a spin may interact with itself (selfish/introvert  individual) even twice (super-selfish/super-introvert individual) and each spin may interact with another one more than once (up to $N$ times in the worst case).

\begin{figure}
\begin{center}
\includegraphics[angle=-90,width=.8\textwidth]{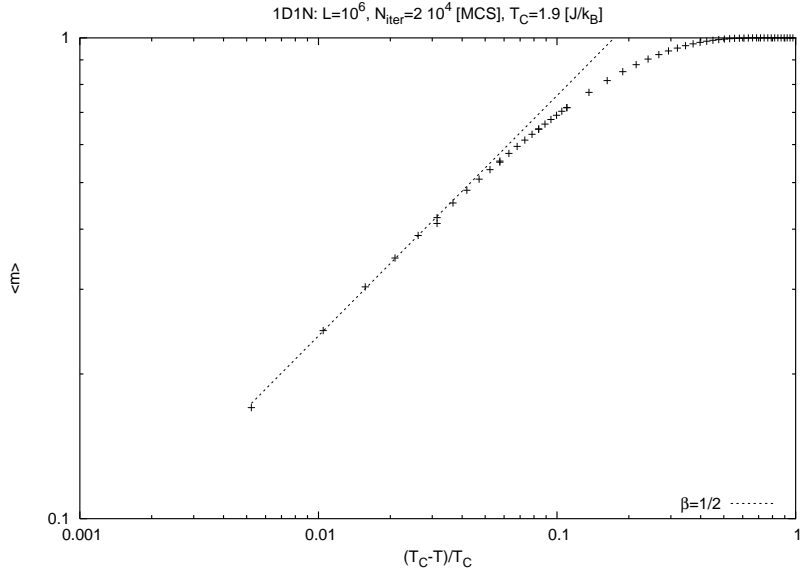}
\end{center}
\caption{The magnetization $\langle m\rangle$ vs. $(T_C-T)/T_C$ for 1D1N case. $T_C\approx 1.9~[J/k_B]$.
The systems contains $N=10^6$ sites.
The simulation takes $2\cdot 10^4$ [MCS].
The dashed line shows $\langle m\rangle \propto ((T_C-T)/T_C)^{1/2}$.}
\label{fig_beta}
\end{figure}

For the 1D1N case --- which corresponds to the P\c{e}kalski model with $p=1$ \cite{pekalski} as well as to Hong-Kim-Choi model \cite{hong}
--- we have found the critical exponent $\beta$ (describing the critical behavior of the magnetization in the vicinity of the transition) equal to $1/2$ (see Fig. \ref{fig_beta}) which is similar to the results of extensive Monte Carlo simulation presented in Ref. \cite{hong}.
We have observed that the average absolute value of the magnetization $\langle |m| \rangle$ for $T\approx T_C$ decreases with the system size $10\le L \le 10^4$, in contrast to Ref. \cite{pekalski}.

\bigskip

{\bf Acknowledgments.} The author is grateful to Dietrich Stauffer for all suggestions and fruitful discussion during his ten-days-long stay in Krak\'ow.
The simulations were carried out in ACK-CYFRONET-AGH.
The machine time on SGI 2800 is financed by the State Committee for Scientific Research (KBN) with grant No. KBN/\-SGI2800/\-AGH/\-???/\-2002.

\end{document}